\input{psfig}
\documentstyle[12pt, epsf]{article}
      \def\cf{\hbox{\it cf.}{}}

\catcode`\@=11

\@addtoreset{equation}{section}
\def\theequation{\arabic{equation}}

\def\@normalsize{\@setsize\normalsize{15pt}\xiipt\@xiipt
\abovedisplayskip 14pt plus3pt minus3pt%
\belowdisplayskip \abovedisplayskip
\abovedisplayshortskip  \z@ plus3pt%
\belowdisplayshortskip  7pt plus3.5pt minus0pt}
\def\small{\@setsize\small{13.6pt}\xipt\@xipt
\abovedisplayskip 13pt plus3pt minus3pt%
\belowdisplayskip \abovedisplayskip
\abovedisplayshortskip  \z@ plus3pt%
\belowdisplayshortskip  7pt plus3.5pt minus0pt
\def\@listi{\parsep 4.5pt plus 2pt minus 1pt
            \itemsep \parsep
            \topsep 9pt plus 3pt minus 3pt}}

\def\underline#1{\relax\ifmmode\@@underline#1\else
        $\@@underline{\hbox{#1}}$\relax\fi}
\@twosidetrue
\relax

\catcode`@=12

\catcode`\@=11

\def\section{\@startsection{section}{1}{\z@}{3.5ex plus 1ex minus
   .2ex}{2.3ex plus .2ex}{\large\bf}}

\def\ps@headings{\def\@oddfoot{}\def\@evenfoot{}
\def\@oddhead{\hbox{}\hfill
        \makebox[.5\textwidth]{\raggedright\ignorespaces --\thepage{}--
        \hfill }}
\def\@evenhead{\@oddhead}
\def\subsectionmark##1{\markboth{##1}{}}
}

\ps@headings

\catcode`\@=12

\relax

\def\r#1{\ignorespaces $^{#1}$}
\def\figcap{\section*{Figure Captions\markboth
        {FIGURECAPTIONS}{FIGURECAPTIONS}}\list
        {Fig. \arabic{enumi}:\hfill}{\settowidth\labelwidth{Fig. 999:}
        \leftmargin\labelwidth
        \advance\leftmargin\labelsep\usecounter{enumi}}}
 \relax
\def\tablecap{\section*{Table Captions\markboth
        {TABLECAPTIONS}{TABLECAPTIONS}}\list
        {Table \arabic{enumi}:\hfill}{\settowidth\labelwidth{Table 999:}
        \leftmargin\labelwidth
        \advance\leftmargin\labelsep\usecounter{enumi}}}
 \relax
\def\reflist{\section*{References\markboth
        {REFLIST}{REFLIST}}\list
        {[\arabic{enumi}]\hfill}{\settowidth\labelwidth{[999]}
        \leftmargin\labelwidth
        \advance\leftmargin\labelsep\usecounter{enumi}}}
 \relax

\catcode`\@=11

\def\marginnote#1{}
\newcount\hour
\newcount\minute
\newtoks\amorpm
\hour=\time\divide\hour by60
\minute=\time{\multiply\hour by60 \global\advance\minute by-
\hour}
\edef\standardtime{{\ifnum\hour<12 \global\amorpm={am}%
    \else\global\amorpm={pm}\advance\hour by-12 \fi
    \ifnum\hour=0 \hour=12 \fi
    \number\hour:\ifnum\minute<100\fi\number\minute\the\amorpm}}
\edef\militarytime{\number\hour:\ifnum\minute<100\fi\number\minute}
\def\draftlabel#1{{\@bsphack\if@filesw {\let\thepage\relax
  \xdef\@gtempa{\write\@auxout{\string
    \newlabel{#1}{{\@currentlabel}{\thepage}}}}}\@gtempa
    \if@nobreak \ifvmode\nobreak\fi\fi\fi\@esphack}
     \gdef\@eqnlabel{#1}}
\def\@eqnlabel{}
\def\@vacuum{}
\def\draftmarginnote#1{\marginpar{\raggedright\scriptsize\tt#1}}
\def\draft{\oddsidemargin -.5truein
        \def\@oddfoot{\sl preliminary draft \hfil
        \rm\thepage\hfil\sl\today\quad\militarytime}
        \let\@evenfoot\@oddfoot \overfullrule 3pt
        \let\label=\draftlabel
        \let\marginnote=\draftmarginnote
   
\def\@eqnnum{(\theequation)\rlap{\kern\marginparsep\tt\@eqnlabel}%
\global\let\@eqnlabel\@vacuum}  }
\def\preprint{\twocolumn\sloppy\flushbottom\parindent 1em
        \leftmargini 2em\leftmarginv .5em\leftmarginvi .5em
        \oddsidemargin -.5in    \evensidemargin -.5in
        \columnsep 15mm \footheight 0pt
        \textwidth 250mmin      \topmargin  -.4in
        \headheight 12pt \topskip .4in
        \textheight 175mm
        \footskip 0pt
        
\def\@oddhead{\thepage\hfil\addtocounter{page}{1}\thepage}
        \let\@evenhead\@oddhead \def\@oddfoot{} \def\@evenfoot{} 
}
\def\titlepage{\@restonecolfalse\if@twocolumn\@restonecoltrue\onecolumn
     \else \newpage \fi \thispagestyle{empty}\c@page\z@
        \def\thefootnote{\fnsymbol{footnote}} }
\def\endtitlepage{\if@restonecol\twocolumn \else  \fi
        \def\thefootnote{\arabic{footnote}}
        \setcounter{footnote}{0}}  
\catcode`@=12
\relax

\def\ps@headings{\def\@oddfoot{}\def\@evenfoot{}
\def\@oddhead{\hbox{}\hfill
        \makebox[.5\textwidth]{\raggedright\ignorespaces --\thepage{}--
        \hfill }}
\def\@evenhead{\@oddhead}
\def\subsectionmark##1{\markboth{##1}{}}
}

\ps@headings

\relax

\def\firstpage#1#2#3#4#5#6{
\begin{document}
\begin{titlepage}
\nopagebreak
\title{\begin{flushright}
        \vspace*{-1.8in}
        {\normalsize CPTH--S473.1096}\\[-9mm]
        {\normalsize LA-UR-96-4027}\\[-9mm]
        {\normalsize hep-th/9611145}\\[4mm]
\end{flushright}
\vfill
{#3}}
\author{\large #4 \\[1.0cm] #5}
\maketitle
\vskip -7mm     
\nopagebreak 
\begin{abstract}
{\noindent #6}
\end{abstract}
\vfill
\begin{flushleft}
\rule{16.1cm}{0.2mm}\\[-3mm]
$^{\star}${\small Research supported in part by the EEC contract
CHRX-CT93-0340.}\\[-3mm] 
$^{\dagger}${\small Laboratoire Propre du CNRS UPR A.0014.}\\
November 1996
\end{flushleft}
\thispagestyle{empty}
\end{titlepage}}

\def\a{\alpha}
\def\b{\beta}
\def\c{\chi}
\def\d{\delta}
\def\e{\epsilon}
\def\f{\phi}
\def\vf{\varphi}
\def\g{\gamma}
\def\h{\eta}
\def\i{\iota}
\def\j{\psi}
\def\k{\kappa}
\def\l{\lambda}
\def\m{\mu}
\def\n{\nu}
\def\p{\pi}
\def\q{\theta}
\def\r{\rho}
\def\s{\sigma}
\def\t{\tau}
\def\u{\upsilon}
\def\x{\xi}
\def\z{\zeta}
\def\D{\Delta}
\def\F{\Phi}
\def\G{\Gamma}
\def\J{\Psi}
\def\L{\Lambda}
\def\O{\Omega}
\def\P{\Pi}
\def\Q{\Theta}
\def\S{\Sigma}
\def\U{\Upsilon}
\def\X{\Xi}
\def\ca{{\cal A}}
\def\cb{{\cal B}}
\def\cc{{\cal C}}
\def\cd{{\cal D}}
\def\ce{{\cal E}}
\def\cf{{\cal F}}
\def\cg{{\cal G}}
\def\ch{{\cal H}}
\def\ci{{\cal I}}
\def\cj{{\cal J}}
\def\ck{{\cal K}}
\def\cl{{\cal L}}
\def\cn{{\cal N}}
\def\co{{\cal O}}
\def\cp{{\cal P}}
\def\cq{{\cal Q}}
\def\car{{\cal R}}
\def\cs{{\cal S}}
\def\ct{{\cal T}}
\def\cu{{\cal U}}
\def\cv{{\cal V}}
\def\cw{{\cal W}}
\def\cx{{\cal X}}
\def\cy{{\cal Y}}
\def\cz{{\cal Z}}

\def\mbox#1#2{\vcenter{\hrule \hbox{\vrule height#2in
                \kern#1in \vrule} \hrule}}  
\def\sq{\,\raise.5pt\hbox{$\mbox{.09}{.09}$}\,}
\def\sqb{\,\raise.5pt\hbox{$\overline{\mbox{.09}{.09}}$}\,}
\def\tri{\triangle}

\def\nabar{{\overline \nabla}}
\def\bR{\overline R}
\def\R{R^2}
\def\Ric{R_{ab}R^{ab}}
\def\Rie{R_{abcd}R^{abcd}}
\def\sqR{\sq R}

\def\nis{\nointerlineskip}
\def\Abar{\vbox{\nis\moveright.33em\vbox{
        \hrule width.35em height.04em}\nis\kern.05em\hbox{$A$}}{}}
\def\Dbar{\vbox{\nis\moveright.20em\vbox{
        \hrule width.50em height.04em}\nis\kern.05em\hbox{$D$}}{}}
\def\Gbar{\vbox{\nis\moveright.20em\vbox{
        \hrule width.50em height.04em}\nis\kern.05em\hbox{$G$}}{}}
\def\mbar{\vbox{\nis\moveright.15em\vbox{
        \hrule width.60em height.04em}\nis\kern.05em\hbox{$m$}}{}}
\def\Rbar{\vbox{\nis\moveright.20em\vbox{
        \hrule width.50em height.04em}\nis\kern.05em\hbox{$R$}}{}}
\def\Vbar{\vbox{\nis\moveright.05em\vbox{
        \hrule width.60em height.04em}\nis\kern.05em\hbox{$V$}}{}}
\def\Xbar{\vbox{\nis\moveright.20em\vbox{
        \hrule width.60em height.04em}\nis\kern.05em\hbox{$X$}}{}}
\def\thetabar{\vbox{\nis\moveright.15em\vbox{
        \hrule width.30em height.04em}\nis\kern.05em\hbox{$\theta$}}{}}
\def\Lambdabar{\vbox{\nis\moveright.25em\vbox{
        \hrule width.35em height.04em}\nis\kern.05em\hbox{${\mit\Lambda}$}}{}}
\def\Sigmabar{\vbox{\nis\moveright.25em\vbox{
        \hrule width.50em height.04em}\nis\kern.05em\hbox{${\mit\Sigma}$}}{}}
\def\phibar{\vbox{\nis\moveright.18em\vbox{
        \hrule width.40em height.04em}\nis\kern.05em\hbox{$\phi$}}{}}
\def\chibar{\vbox{\nis\moveright.12em\vbox{
        \hrule width.40em height.04em}\nis\kern.05em\hbox{$\chi$}}{}}
\def\psibar{\vbox{\nis\moveright.23em\vbox{
        \hrule width.40em height.04em}\nis\kern.05em\hbox{$\psi$}}{}}
\def\debar{\vbox{\nis\moveright.18em\vbox{
        \hrule width.35em height.04em}\nis\kern.05em\hbox{$\partial$}}{}}
\def\delbar{\vbox{\nis\moveright.10em\vbox{
        \hrule width.63em height.04em}\nis\kern.05em\hbox{$\nabla$}}{}}
\def\delhat{\HAt\de}
\def\Dbarhat{\cap\Dbar}
\def\delbarhat{\Cap\delbar}

\def\Rarr{\rightarrow}
\def\rarr{\rightarrow}
\def\Larr{\leftarrow}
\def\larr{\leftarrow}

\def\tr{\rm{Tr}}
\def\trp{\rm{Tr'}}
\def\Y{Y_{_{JM}}}
\def\Ys{Y^*_{_{JM}}}
\def\Yh{Y_{\ha M}}
\def\dT{{\dot{T}}_{JM}^{(+)}}
\def\phd{\varphi^{\dag}_{_{JM}}}
\def\jo{J_1}
\def\jt{J_2}
\def\mo{M_1}
\def\mt{M_2}
\def\pa{\partial}
\def\de{\nabla}
\def\Ldag{L^{\dagger}}

\def\su{\sum}
\def\iff{\leftrightarrow}
\def\conj{{\hbox{\large *}}}
\def\ltap{\raisebox{-.4ex}{\rlap{$\sim$}} \raisebox{.4ex}{$<$}}
\def\gtap{\raisebox{-.4ex}{\rlap{$\sim$}} \raisebox{.4ex}{$>$}}
\def\TH{{\raise.2ex\hbox{$\displaystyle \bigodot$}\mskip-4.7mu \llap H \;}}
\def\face{{\raise.2ex\hbox{$\displaystyle \bigodot$}\mskip-2.2mu \llap {$\ddot
        \smile$}}}

\def\simlt{\stackrel{<}{{}_\sim}}
\def\simgt{\stackrel{>}{{}_\sim}}
\newcommand{\dal}{\raisebox{0.085cm}
{\fbox{\rule{0cm}{0.07cm}\,}}}
\newcommand{\dt}{\partial_{\langle T\rangle}}
\newcommand{\dtbar}{\partial_{\langle\bar{T}\rangle}}
\newcommand{\al}{\alpha^{\prime}}
\newcommand{\mst}{M_{\scriptscriptstyle \!S}}
\newcommand{\mpl}{M_{\scriptscriptstyle \!P}}
\newcommand{\dv}{\int{\rm d}^4x\sqrt{g}}
\newcommand{\lv}{\left\langle}
\newcommand{\rv}{\right\rangle}
\newcommand{\ph}{\varphi}
\newcommand{\abar}{\bar{a}}
\newcommand{\sbar}{\,\bar{\! S}}
\newcommand{\xbar}{\,\bar{\! X}}
\newcommand{\fbar}{\,\bar{\! F}}
\newcommand{\zbar}{\bar{z}}
\newcommand{\dbar}{\,\bar{\!\partial}}
\newcommand{\tbar}{\bar{T}}
\newcommand{\taubar}{\bar{\tau}}
\newcommand{\ubar}{\bar{U}}
\newcommand{\ybar}{\bar{Y}}
\newcommand{\phb}{\bar{\varphi}}
\newcommand{\cm}{Commun.\ Math.\ Phys.~}
\newcommand{\prl}{Phys.\ Rev.\ Lett.~}
\newcommand{\pr}{Phys.\ Rev.\ D~}
\newcommand{\pl}{Phys.\ Lett.\ B~}
\newcommand{\ibar}{\bar{\imath}}
\newcommand{\jbar}{\bar{\jmath}}
\newcommand{\np}{Nucl.\ Phys.\ B~}
\newcommand{\A}{{\cal A}}
\newcommand{\be}{\begin{equation}}
\newcommand{\en}{\end{equation}}
\newcommand{\gsi}{\,\raisebox{-0.13cm}{$\stackrel{\textstyle
>}{\textstyle\sim}$}\,}
\newcommand{\lsi}{\,\raisebox{-0.13cm}{$\stackrel{\textstyle
<}{\textstyle\sim}$}\,}
\date{}
\firstpage{3118}{IC/95/34}
{\large\bf  Criticality and Scaling in 4D Quantum Gravity$^{\star}$}  
{Ignatios Antoniadis$^{\,a}$, Pawel O. Mazur$^{\,b}$ and Emil
Mottola$^{\,c}$}
{\normalsize\sl
$^a$Centre de Physique Th\'eorique, Ecole Polytechnique,$^\dagger$
{}F-91128 Palaiseau, France\\[-3mm]
\normalsize\sl
$^b$Dept. of Physics and Astronomy, University of S. Carolina, Columbia, SC
29208 USA\\[-3mm]
\normalsize\sl
$^c$Theor. Division, T-8, M.S. B285, Los Alamos National Laboratory,
NM 87545 USA.}  
{We present a simple argument which determines the critical value of
the anomaly coefficient in four dimensional conformal factor
quantum gravity, at which a phase transition between a smooth
and elongated phase should occur. The argument is based on the contribution of
singular configurations (``spikes") which dominate the partition function
in the infrared. The critical value is the analog of $c=1$ in the 
theory of random surfaces, and the phase transition is similar 
to the Berezenskii-Kosterlitz-Thouless transition. 
The critical value we obtain is in agreement with the 
previous canonical analysis of physical states of the conformal 
factor and may explain why a smooth phase of quantum gravity has not yet
been observed in simplicial simulations. We also rederive the 
scaling relations in the smooth phase in light of this determination of
the critical coupling.}

In recent work we have developed a framework which permits the study of
the effects of gravitational fluctuations at large distances
\cite{am}-\cite{states}. It is based on the quantization of a low energy
effective action for the spin-$0$ or conformal part of the metric, which is
determined by the  trace anomaly. This action implies the existence of an
infrared stable fixed point for quantum gravity in four dimensions with
nontrivial scaling behavior \cite{am}. On the other hand, quantum gravity
can be reformulated as a statistical model on a random lattice and
studied  numerically \cite{simplicial}. There is then the attractive
possibility for testing the continuum predictions of an infrared fixed
point by quantitative measurements in the lattice approach \cite{scaling,aj}.

In a previous letter \cite{scaling} we discussed the scaling relations for
the partition function and observables in the conformal phase, based on
the results of the infrared fixed point behavior found earlier.
Since that time we have performed an extensive canonical analysis of the
constraints of diffeomorphism invariance on the Einstein universe $R\times
S^3$, and found a discrete spectrum
of physical states in the Fock space which survive the imposition of
the constraints \cite{states}. Each of these physical states corresponds
to an operator of scaling dimension or conformal weight $4$ constructed
from the conformal part of the metric and its derivatives.  
The existence of these states in a pure gravity theory without matter is in
sharp contrast to the analogous quantization
of the Liouville theory on the cylinder $R\times S^1$ where there is
only one state (the ``vacuum") and only one operator (the identity or
volume operator). In four dimensions, there is a tower of operators
describing marginal deformations of the the theory away from its
infrared fixed point. In the semi-classical limit, the two lowest of 
these become the Einstein-Hilbert and cosmological (or volume) terms. 
Each of these two operators comes with its own scaling behavior and anomalous
dimensions, and each of them seems to imply a different value of
the coupling where a phase transition to a highly nonclassical
(or branched polymer) phase of quantum gravity could occur. 
Since the situation is more complicated than the $D=2$ case, 
the scaling behavior of observables should be reconsidered from the 
present vantage point of a more complete understanding of the physical 
states and the operators that create them, and the critical coupling 
determined by the point at which the scaling exponents first cease to be real. 

In two dimensions there is a simple argument for the critical value $c=1$
of the embedding dimension in Polyakov's theory of random surfaces (or
noncritical bosonic string theory) \cite{cates}. The argument is
reminiscient of the Berezenskii-Kosterlitz-Thouless (BKT) argument for a
phase transition in the 2D x-y model \cite{bkt}, in that it relies upon
constructing a singular solution to the classical equations (called a
``spike" in this context), which makes a contribution to the  action that
depends logarithmically on the infrared cut-off.  Since the entropy of such
configurations also grows logarithmically with the volume of the system, 
there is a critical value of the coupling at which the entropy always
overwhelms the action and the partition function is dominated by a dense
gas of such singular spikes. Conversely, if the coupling is adjusted in the
opposite direction then such configurations are always suppressed in the
infinite volume limit, and the geometry of the surface can be reasonably
smooth. Since the coupling is also the coefficient of the trace anomaly in
2D, this argument also tells us what the critical value of the matter
central charge is, and why one should expect a phase transition from a
smooth to a branched polymer phase at this critical coupling of $c=1$. This
phase transition has been verified  in the dynamical triangulation approach
to 2D random surfaces \cite{c=1}.

Since a branched polymer (or elongated) phase has been observed also in 4D
simplicial quantum gravity \cite{simplicial,aj}, it is quite natural to
suspect that an analogous argument involving the ``liberation of spikes" in the
subcritical region of a coupling constant ($Q^2$ defined below) should
apply again in this case, as remarked in a recent paper \cite{kj}.  
In this note we develop this idea that the  
``half-wormhole" or ``spike" configurations are the relevant ones
for the phase transition at a certain critical value of the anomaly
coefficient and we study the implications for 4D simplicial simulations 
and scaling relations in the smooth phase. 

{\em The Critical Coupling}.
In four dimensions, the analog of the Polyakov-Liouville theory of random
surfaces is obtained by considering the effective action for the conformal 
factor of the metric induced by the 4D trace anomaly. This
effective action takes the form \cite{am}
\be
S_E[\sigma] = {Q^2 \over (4 \pi)^2}\int d^4 x \sqrt{g} \bigl[
\s  {\D}_4 \s + \hbox{$1\over 2$}\bigl(G - \hbox{$2 \over 3$} \sq R \bigr) \s
\bigr]\ , 
\label{act}
\en
where ${\D}_4$ is the unique Weyl covariant fourth order differential
operator on
scalars, $e^{2\sigma}$ is the conformal factor of the metric (taken here with
Euclidean signature), and $Q^2$ is the coefficient of the Gauss-Bonnet term
$G$ in the trace anomaly. It is normalized such that \cite{amm}
\be
Q^2 = {1 \over 180}(N_S + \hbox{$11\over 2$} N_{WF} + 62 N_V - 28) + 
Q^2_{grav}\ ,
\label{cent}
\en
where $N_S, N_{WF}, N_V$ are the number of free scalars, Weyl fermions and
vector fields and $Q^2_{grav}$ is the contribution of spin-$2$ gravitons,
which has not yet been determined unambiguously. The $-28$ contribution
is that of the $\sigma$ field itself which is the only known negative 
contribution to $Q^2$. Thus, it is $Q^2$ which plays the role analogous
to matter central charge and $\Delta_4$ which plays the role of
the kinetic operator $\sq$ in $D=2$. In flat background coordinates 
$\Delta_4 = \sq^2$.

Now we make the following observation. Because of the fourth order 
conformal differential operator matched to the number of dimensions, the
propagator of this operator is a logarithm, just as is that of $\sq$ in
$D=2$. This means that we have a situation analogous to that in 2D.
Although the Mermin-Wagner theorem forbids a spontaneously broken
phase with massless excitations, the 2D x-y model does exhibit a BKT phase
transition at a certain critical temperature ({\it i.e.} coupling in the 2D
Euclidean field theory), which is just the result of the logarithmic growth
of the massless conformal invariant $\sq^{-1}$ propagator in $D=2$. Because
the conformal propagator $\Delta_4^{-1}$  in four dimensions has logarithmic
growth at large distances, one should expect a BKT-like phase transition at
a certain critical coupling $Q^2_{cr}$ in 4D for essentially the same reason
as in the original x-y model or in the 2D theory of random surfaces. 

Indeed, we can construct exactly the same spike solution in $D=4$ as in
$D=2$,
\be
\sigma_S (x) = q \ln |x-x_0|
\en
with exactly the same interpretation. It is the solution of the classical
equation
following from (\ref{act}) with a delta function singularity at $x=x_0$ of
strength $q$.  In order to regulate the singular behavior of this configuration
one may introduce an ultraviolet (UV) cut-off $a$, replacing $|x-x_0|^2$ by
$1 + |x-x_0|^2/a^2$ in the logarithm.  This UV cut-off will be of order of
the lattice spacing in the simplicial simulations. The infrared (IR) 
cut-off of the logarithm will be provided by the finite volume.   

Evaluating the action $S_E$ on this spike configuration in a large but finite 
spherical volume of radius $L$ we find
\be
S_E[\sigma_S] = {1\over 2} Q^2 q^2 \ln \left({L\over a}\right)\ .
\en
On the other hand the number of ways of placing this configuration in the
volume $V$ is proportional to the volume, $V \sim L^4$. Hence, the entropy grows
like $4 \ln ({L\over a})$ and the free energy of such configurations
behaves like
\be
{}F [\sigma_S] =  \left({1\over 2} Q^2 q^2 - 4\right) \ln \left({L\over
a}\right)
\label{ener}
\en
for large $L$. It follows that for $Q^2 > 8/q^2$  these singular configurations
will have positive free energy and be suppressed in the thermodynamic limit
$L \rightarrow \infty$, while if the inequality is reversed we must expect
them to dominate the partition function.
Consequently, there is a critical value of the coupling $Q^2$ at which we expect
to see a phase transition from a phase with many sharp elongated spikes to one
where such singular configurations are suppressed.
The analogy to the BKT argument involving vortices in the
x-y phase field is obvious.

In two dimensions, Cates argued that the value $q = -1$ (in the present
notation) is special because this is the strength at which the singularity
is first strong enough to give a divergent contribution to the 
classical volume (area in $D=2$) as the cut-off $a$ is removed \cite{cates}. Hence, these are the configurations with the lowest free energy that can dominate the partition function in the continuum limit $L \gg a$. 
Since the volume or cosmological operator is a marginal deformation of 
the free Liouville theory, this argument for $q=-1$ can be turned into 
a renormalization group analysis in the IR as well, by 
taking into account the gravitational ``dressing" of the
volume operator due to the loop corrections of the free $\sq^{-1}$
propagator near its Gaussian fixed point. Then we find that $q=-1$ is
precisely the condition that justifies the neglect of the cosmological term and
the use of the  spike solution to the free Liouville theory in the
infinite volume limit $L \rightarrow \infty$. Let us give
this argument why $q=-1$ is the relevant spike solution in four dimensions
as well.

In four dimensions, there are two terms with fewer derivatives
that can be added to the free action (\ref{act}).
They are the volume or cosmological term,
\be
\lambda S_0 [\sigma] = \lambda \int d^4 x\,\sqrt{g} = \lambda \int d^4 x\, e^{4\sigma} \rightarrow 
\lambda \int d^4 x\, e^{\beta_0\sigma}\,.
\label{Cos}
\en
and the Einstein-Hilbert term,
\begin{eqnarray}
{1\over 2\kappa} S_2 [\sigma] &=& -{1\over 2\kappa}\int d^4 x\,\sqrt{g}\, R = {3\over \kappa} \int d^4x\,e^{2\sigma}
\left[\sq\sigma + (\partial\sigma)^2\right]\nonumber\\
&\rightarrow& -{3\over 4\kappa}\int d^4x\,\left[ \beta_2^2(\partial\sigma)^2
e^{\beta_2\sigma} - {72\pi^2\over \kappa} f(Q^2) e^{2\beta_2\sigma}\right]\,.
\label{Ein}
\end{eqnarray}
The integrands in these expressions have engineering dimensions $0$ and $2$
respectively, and so they must be multiplied by powers of $e^{\sigma}$ to
give a scalar density with total conformal weight $4$. In the last forms of
(\ref{Cos}) and (\ref{Ein}) we have allowed  for the possibility that these
terms are gravitationally dressed and that the classical scaling
codimensions $(\beta_0)_{cl} = 4$ and $(\beta_2)_{cl} = 2$ are modified at
the quantum level. {}For the Einstein term involving two derivatives
of the $\sigma$ field there is also the possibility of renormalization
group mixing with operators of lower dimension, which is represented by the
$f(Q^2)$ term in eq.~(\ref{Ein}). Indeed, a covariant computation of the
renormalization of these operators due to the loop effects of the free
$\sq^2$ propagator shows that $\beta_0$ and $\beta_2$ satisfy the equations,
\be
\beta_0 = 4 + {\beta^2_0 \over 2 Q^2}\,,
\label{cosscal}
\en
and
\be
\beta_2 = 2 + {\beta_2^2 \over 2Q^2}\,,
\label{einscal}
\en
while the operator mixing is given by
\be
f(Q^2) = {\alpha^2 \over Q^2} \left[ 1 + {4\alpha^2\over Q^2}
+ {6\alpha^4\over Q^4}\right]\,,
\label{mix}
\en
and $\alpha \equiv {\beta_2/ 2}$ in the present notation \cite{am}.  
The classical values for the codimensions $\beta_0$ and $\beta_2$ are
recovered only in the limit $Q^2 \rightarrow \infty$, where quantum
fluctuations of the conformal factor are suppressed. Notice that
this classical limit is the analog of the opposite limit of the central
charge ({\it i.e.} $c\rightarrow -\infty$) in the 2D case, since $Q^2$ is
positive for free conformal matter fields. 

Corroboration of the anomalous scaling dimensions (\ref{cosscal}) and
(\ref{einscal}) comes from the canonical quantization of the $\sigma$
theory on the Einstein space $R\times S^3$ where an infinite tower of
discrete diffeomorphic invariant states, labelled by the integers, was
obtained \cite{states}. In the semi-classical limit $Q^2 \rightarrow
\infty$, these states are created by operators which are volume integrals
of integer powers of the Ricci scalar, {\it i.e.},
\be
\int d^4 x\,\sqrt{g} R^n \sim \int d^4 x\, e^{\beta_{2n}\sigma}
\left[(\partial\sigma)^{2n} +\cdots\right]\,,
\label{nops}
\en
where
\be
\beta_{2n} = 4-2n + {\beta_{2n}^2 \over 2Q^2}\,.
\en
The ellipsis in (\ref{nops}) refers to the operators with lower numbers of
derivatives which mix with $(\partial\sigma)^{2n}$ under the renormalization
group at finite $Q^2$, analogous to (\ref{Ein}) and (\ref{mix}) for $n=1$. The
exact forms of this operator mixing is not determined by the canonical analysis,
but  the values of all the $\beta_{2n}$ are fixed by the Hamiltonian state 
condition (the analog of the $L_0-1$ condition on the physical states in $D=2$).
The first two of the discrete physical states  are created by operators of the
form (\ref{Cos}) and (\ref{Ein}) and the values of the scaling exponents are
identical to (\ref{cosscal}) and (\ref{einscal}) obtained in the covariant Euclidean approach. 
 
Now the volume term evaluated on the spike
configuration behaves like
\be
S_0 [\sigma_S] \sim -{a^4 \over (4 + \beta_0 q)} 
\left[c_0 -\left({L\over a}\right)^{4 + \beta_0 q}\right]
= {2 a^4Q^2\over \beta_0^2}\left[c_0 - \left({L\over a}\right)^{-{\beta_0^2\over 2 Q^2}}\right]\,,
\en
for $L \gg a$ and $q=-1$, which is the only value for which the integral
is IR convergent for any positive value of $Q^2$ (for which Re $\beta_0 > 4$). In this expression $c_0$ is a positive constant of order unity whose value depends on the precise way the UV cut-off $a$ is introduced. In a similar manner, the Einstein term evaluated on the spike 
configuration behaves like
\be
S_2 [\sigma_S] \sim  {q^2 a^2 \over (2 + \beta_2 q)}
\left[c_2 - \left({L\over a}\right)^{2 + \beta_2 q} 
\right] = -{2a^2Q^2\over \beta_2^2}
\left[c_2 -
\left({L\over a}\right)^{-{\beta_2^2\over 2 Q^2}}\right]
\en
for $L \gg a$ and $q= -1$, where $c_2$ is another constant (dependent
on $a^2/\kappa$). 
Inspection of the previous form shows that $q=-1$ is again the only
value of $q$ for which the Einstein action evaluated on the spike
configuration is convergent as $L\rightarrow \infty$, for any positive value of $Q^2$ (for which Re $\beta_2 > 2$).

Hence, knowledge of the scaling exponents of the Einstein and volume operators
allows us to compute their values on the singular spike configurations
and show that they are subdominant to $S_E$ in the infinite volume continuum
limit for all sufficiently large and positive $Q^2$, if and only if $q=-1$.
With $q=-1$ selected in this way, we obtain from our previous evaluation
of the free energy of the spike (\ref{ener}) the critical value,
\be
Q^2_{cr} = 8\,.
\en     
{}For $Q^2 > 8$ the solution of the quadratic relations (\ref{einscal}) and
(\ref{cosscal}), {\it viz.},
\be
\beta_0 = Q^2 \left( 1 - \sqrt{1 - {8\over Q^2}}\right)
\label{beta0}
\en
and
\be
\beta_2 = Q^2 \left( 1 - \sqrt{1 - {4\over Q^2}}\right)
\label{beta2}
\en
are indeed both real and Re $\beta_0 > 4$, Re $\beta_2 > 2$ for
$8<Q^2<\infty$, so that the previous argument for the irrelevance of the
Einstein and cosmological terms in the infinite volume limit are justified
{\it a posteriori}. We have chosen the minus sign for the square roots in
(\ref{beta0}) and (\ref{beta2}) in order that $\beta_{2n} \rightarrow
(\beta_{2n})_{cl}=4-2n$ in the semi-classical limit $Q^2 \rightarrow
\infty$. 

When $Q^2 < 8$, then  one of these exponents (namely $\beta_0$) becomes
complex and we should expect qualitatively different behavior of the theory.
This is an independent indication of the critical value $Q^2_{cr} = 8$.
{}From the simple BKT-like argument, we expect that
$Q^2 < 8$ corresponds to a phase which is dominated by a dense gas of $q=-1$
spikes. The metric of the singular $q=-1$ configuration
(with the origin, $x_0 =0$) is
\be
ds^2 = e^{2\sigma_S}d\bar s^2 ={ (dr^2 + r^2 d\Omega^2)\over r^2} = 
(d \ln r)^2 + d\Omega^2\ ,
\en
which is just that of $R\times S^3$, if we start with the flat $R^4$ metric
background, $d\bar s^2$. Since the $S^3$ has arbitrary radius and the $R$ is the
entire real line, this geometry is arbitrarily elongated and thin. Since any
region is locally like that of $R^4$, a ``gas" of such configurations means that
these ``spikes" (or perhaps more appropriately, ``tubes" or ``punctures") will
look like a large number of branches with a hierarchy of thinner and thinner
sub-branches. A random geometry with many such spiky extrusions looks very much
like the elongated phase described in ref.~\cite{aj}. Conversely, if $Q^2 > 8$
then the scaling dimensions (\ref{beta0}) and (\ref{beta2}) are real, 
the spikes are suppressed and one would
expect to find that the partition function is dominated by much smoother
configurations, which at very large $Q^2$ are well-approximated by
semi-classical metrics.

Thus, $Q^2_{cr} =8$ behaves in many respects like the $c=1$ case in 2D
gravity, with the important difference that  the addition of conformal
matter fields brings us into the smooth phase $Q^2 > 8$ rather than into the
elongated or branched polymer phase $c>1$ (see eq.~(\ref{cent})). Now in
pure simplicial gravity, {\it i.e.} without the introduction of matter,
only an elongated phase and a crumpled phase (where the geometries collapse
upon themselves) have been found. There is a transition between
them as the lattice coupling corresponding to the Einstein term is varied,
although apparently there is as yet no universal agreement as to the order of 
this transition.
{}From the point of view of the considerations presented here the numerical
results are consistent with the hypothesis that pure gravity has $Q^2 =
Q^2_{grav} \simlt 8$, and only a spiky elongated phase in the continuum
limit. We should emphasize in this connection that the anomaly generated
action $S_E$ is positive definite for $Q^2 > 0$ so that there is no
conformal factor problem, nor any need for a conformal rotation as in the
Einstein theory. Hence, the statistical continuum theory of random 
four-geometries described by (\ref{act}) is well-defined for $0< Q^2 < 8$
even if these geometries turn out to be very jagged.

The main theoretical difficulty in determining $Q^2_{grav}$ is that the
Einstein theory is neither conformally invariant nor free, so that
a method for evaluating the strong infrared effects of spin-$2$
gravitons in the continuum must be found that is insensitive to 
ultraviolet physics. In ref.~\cite{amm} we performed a strictly
perturbative  computation which gives the value $Q^2_{grav}=1411/180
\approx 7.9$ for the graviton contribution. Since the method used is based
on the heat kernel expansion, there is the possibility of mixing up
ultraviolet with infrared effects in this evaluation, and we 
cannot regard it as definitive. However, a computation using the
totally different conformally invariant Weyl tensor-squared action 
leads to a similar value $Q^2_{grav}=8.7$. Hence it is likely that the
correct infrared graviton contribution to $Q^2$ is in the neighborhood of
$8$, and that most of this large value is due to the spin of the
field, which is larger than any other individual field's contribution of
lower spin. If the value of $Q^2$ in the pure gravity theory
turns out to be less than eight, then one would not
expect to find a second order phase transition to the continuum limit in the
simplicial simulations with geometries that are smooth. Instead the
continuum limit
of the random geometries would be an elongated phase, filled with many
spiky extrusions.

If this is indeed the correct interpretation of the numerical results to
date, then the introduction of some number of free conformal matter fields
to the simplicial simulations would push the value of $Q^2$ above $8$ and
lead to a continuum limit exhibiting a smooth phase, perhaps by softening
a weakly first order transition between the elongated and crumpled
phases into a second order transition with the new phase appearing in the
phase diagram extended into  the $Q^2> Q^2_{grav}$ direction. Taking the
$1411/180$ number at face value and using the known contributions to $Q^2$
for lower spin fields in eq.~(\ref{cent}), one finds that it would take the
introduction of $57$ conformal scalars, but {\it only one photon} to induce
this transition to the smooth phase.

{\em Scaling relations}.
Assuming that the smooth phase with $Q^2 > 8$ exists, the exponents make
definite predictions for the scaling of the fixed volume partition function,
\be
Z(\kappa, \lambda) \equiv \int [{\cal D}\sigma]\exp \left( - S_E[\sigma]
-{1\over 2\kappa}S_2[\sigma] -\lambda S_0[\sigma]\right)\,.
\en
The covariant continuum measure $[{\cal D}\sigma]$ has been discussed
in several previous articles \cite{amm,measure}. By inserting a delta
function of physical four-volume, we obtain the fixed volume partition
function
\be
Z(\kappa ; V) \equiv \int [{\cal D}\sigma]\exp \left( - S_E[\sigma]
-{1\over 2\kappa}S_2[\sigma]\right) \delta\left( S_0[\sigma] - V\right)\,.
\en

If the effective action appearing here, namely the sum of the Einstein
and anomaly induced terms, is written in an arbitrary curved background and
one studies the effect on the finite volume partition function of a shift
in $\sigma$,
\be
\sigma \rightarrow \sigma + \omega
\label{shift}
\en
then one finds that
\be
Z(\kappa ; V) = \exp\left[-\omega\left(\beta_0 + Q^2\chi_{_E}\right)\right]
Z(\kappa e^{-\omega \beta_2}; e^{-\omega\beta_0 }V)\,,
\label{Zshift}
\en
where the Euler number $\chi_{_E} = 2$ for fixed $S^4$ topology.
Since $\sigma$ has been integrated out, the resulting $Z(\kappa ; V)$ must
be independent of the shift (\ref{shift}), which is only consistent with
(\ref{Zshift}) if the fixed volume partition function can be expressed
in the form,
\be
Z(\kappa ; V) = V^{-1- 2{Q^2 \over \beta_0}}\ 
\tilde Z(\kappa V^{- {\beta_2\over \beta_0}})\,,
\label{kap}
\en
for some function $\tilde Z$ of a single argument. This scaling relation
differs from that obtained in eq.~(10) in our previous letter
\cite{scaling}, since now we have allowed the volume and Einstein terms
(\ref{Cos}) and (\ref{Ein}) to have independent scaling exponents
in agreement with our subsequent canonical analysis \cite{states} and ref.
\cite{s}.

By using the relation (\ref{beta0}) we now obtain the susceptibility exponent
\be
\gamma (Q^2) = 2 - 2 {Q^2\over\beta_0} = -2\, {\sqrt{1 - {8\over Q^2}}
\over 1 - \sqrt{1- {8\over Q^2}}}
\label{sus}
\en
instead of eq.~(23) of ref.~\cite{scaling}. Notice that if $Q^2$ is close to
$8$ then  the susceptibility exponent is close to zero, which is what has
been found in the simulations to date. As shown in Fig.~1, $\gamma$ is
negative for $Q^2>8$ and approaches zero from below as $Q^2\to 8$. 

The form (\ref{kap}) also implies that we should require $\kappa V^{-
{\beta_2\over\beta_0}}$ to remain finite in the scaling limit, or
equivalently,
\be
\kappa \sim V^{\delta} \qquad {\rm as } \qquad V\rightarrow \infty\,,
\en
with
\be
\delta = {\beta_2\over \beta_0} = {1 - \sqrt{1-{4\over Q^2}}\over
1- \sqrt{1 - {8\over Q^2}}}\,.
\label{del}
\en

\begin{figure*}
\vspace{-3cm}
\hspace*{-1.5cm}
\psfig{figure=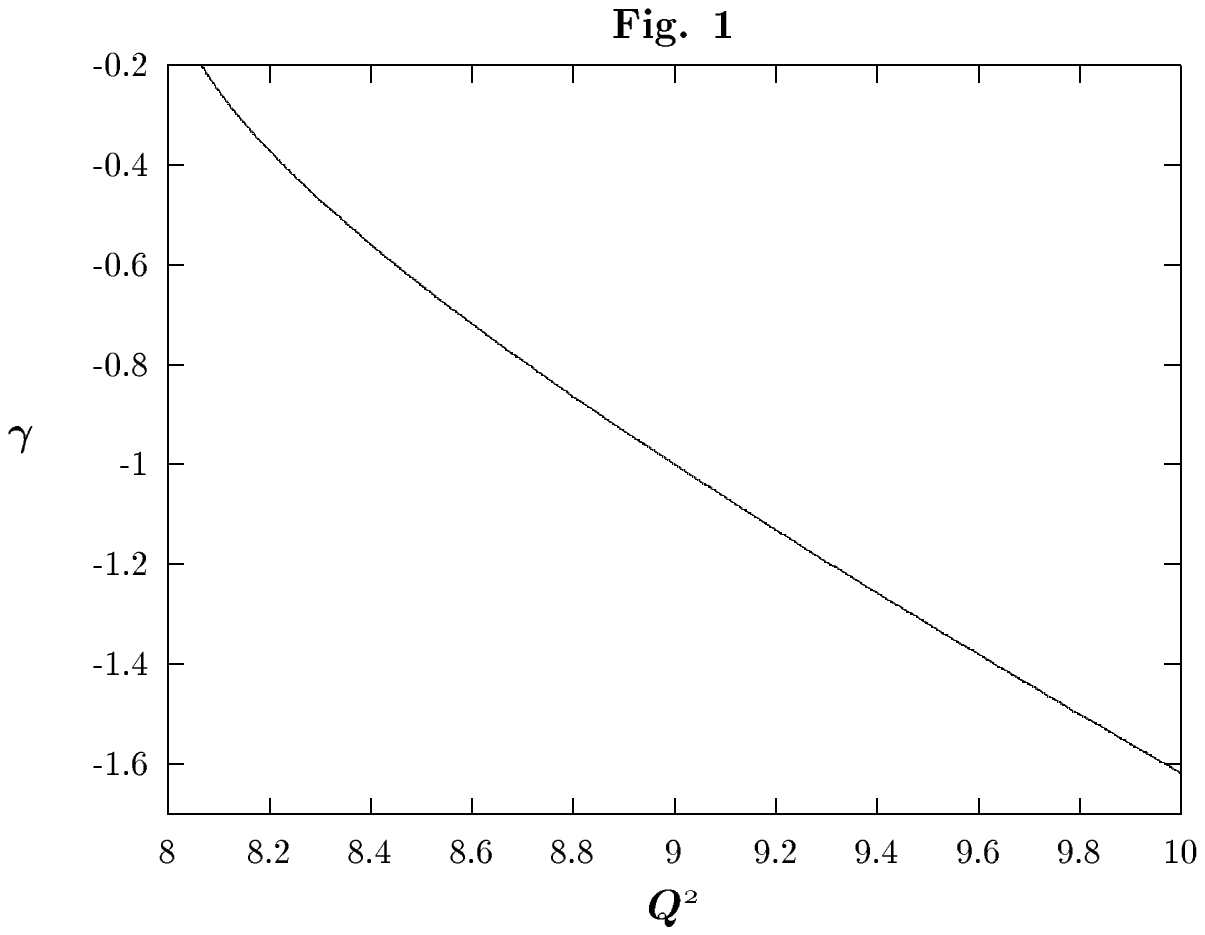,height=18cm,width=9.2cm}
\hspace*{-0.82cm}
\psfig{figure=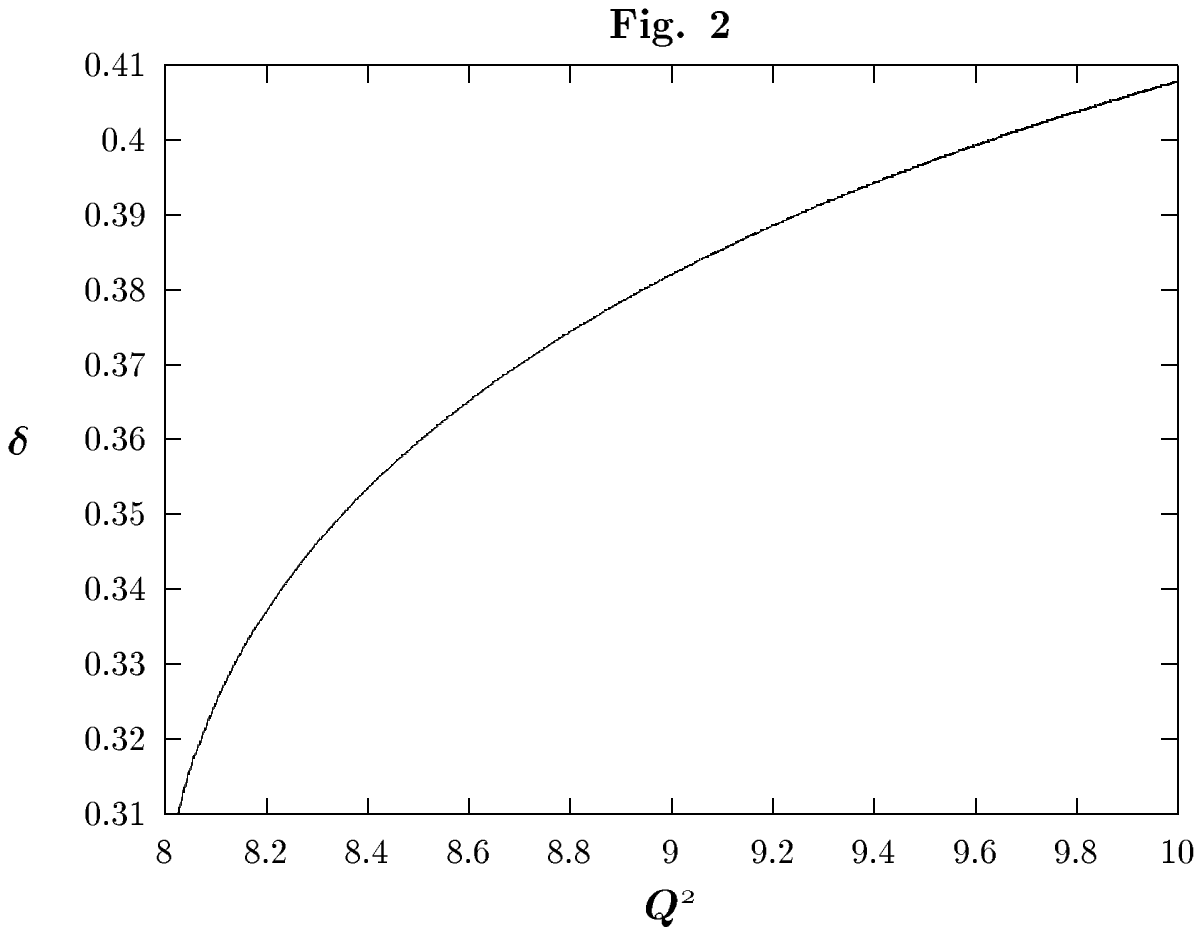,height=18cm,width=9.1cm}
\vspace*{-9.2cm}

Figures 1 and 2. The susceptibility and scaling exponents $\gamma$
and $\delta$ as functions of the anomaly coefficient $Q^2$.
\vspace*{0.2cm}
\end{figure*}

We remark that this relation differs from eq.~(22) of ref.~\cite{scaling}
because of the addition of the second independent cosmological
operator which we did not consider in that work. The exponent $\delta$ is
plotted as a function of $Q^2$ in Fig.~2. 
{}For $Q^2$ near $8$, $\delta \simeq 0.3$, which differs from the value
$0.47\pm 0.03$ reported in the simulations of ref.~\cite{aj}. However, if our
hypothesis is correct, namely that $Q^2$ for pure gravity is less than $8$,
then the smooth phase where eq.~(\ref{del}) applies has not yet been produced
in the simulations. Of course, when the numerical situation has been
clarified and an unambiguous continuum limit with a smooth phase 
has been demonstrated, then there is only one free parameter and 
both (\ref{sus}) and (\ref{del}) must be consistent with the same value 
for $Q^2$, if the conformal fixed point relations predicted by the 
continuum theory are correct.

{}Finally, one can consider any operator composed of either metric or
matter fields with scaling dimension $\bar\Delta$, in the absence of
dressing. By inserting this operator in the fixed volume partition function
\be
\langle {\cal O}_{\bar\Delta} \rangle_{_V} = {1\over Z(\kappa ;V)}
 \int [{\cal D}\sigma]\,{\cal O}_{\bar\Delta}\,\exp \left( - S_E[\sigma]
-{1\over 2\kappa}S_2[\sigma]\right) 
 \delta\left( S_0[\sigma] - V\right)\,,
\en
and repeating the shift (\ref{shift}) and scaling argument
for this quantity, we find
\be
\langle {\cal O}_{\bar\Delta} \rangle_{_V} \sim V^{1 - {\Delta\over 4}}\ ,
\en
where the full scaling dimension $\Delta$ is related to the ``classical"
dimension ${\bar\Delta}$ by
\be
\Delta = 4 - 4\, {\beta_{\bar\Delta}\over \beta_0} = 
4\,{ \sqrt{1 - 2 {(4-\bar\Delta)\over Q^2}} - \sqrt{1 - {8\over Q^2}}\over
1 - \sqrt{1 - {8\over Q^2}}}\ .
\label{dress}
\en
Here $\beta_{\bar\Delta}$ is the relevant codimension of the
gravitational dressing determined by
\be
\beta_{\bar\Delta} =4 - \bar\Delta + {\beta^2_{\bar\Delta} \over 2 Q^2}\,.
\label{cod}
\en
At large $Q^2$ one has:
\be
\Delta-{\bar\Delta}={\beta_{\bar\Delta}^2\over 32Q^2}\Delta (4-\Delta)=
{1\over 2Q^2}{\bar\Delta}(4-{\bar\Delta})+\cdots
\en
{}For the volume and Einstein operators for which $\bar\Delta$ is $0$ and
$2$, respectively, we recover the previous relations (\ref{cosscal}) and
(\ref{einscal}).

If these scaling relations can be verified in the smooth phase, and the infrared conformal fixed point predicted by the action (\ref{act}) confirmed,
then the lattice could provide a nonperturbative method
for measuring the contribution $Q^2_{grav}$. In addition to being
interesting in its own right by exploring a nontrivial fixed point
of 4D quantum gravity, the scaling relations could also find
applications in cosmology \cite{sky}.   

\noindent{\bf Acknowledgments} 

We are grateful to J. Jurkiewicz and A. Krzywicki for stimulating discussions
in developing this idea and to J. Ambj{\o}rn and J. Smit for useful conversations. P. O. M. and E. M. thank the Centre de Physique
Th\'eorique at the Ecole Polytechnique for its hospitality. Finally, we would
like to thank S. Lola for helping us to produce the figures.
     

\end{document}